\documentstyle[12pt,epsf]{ioplppt}
\begin{document}
\pagestyle{empty}
\jl{1}
\eqnobysec
\def\qua{{
        \setlength{\unitlength}{0.1mm}
        \begin{picture}(24,20)(-2,1)
        \put(0,0){\line(1,0){20}}
        \put(0,0){\line(0,1){20}}
        \put(20,20){\line(-1,0){20}}
        \put(20,20){\line(0,-1){20}}
        \end{picture}
        }}
\newcommand{\av}[1]{\langle #1 \rangle}
\renewcommand{\e}[1]{e^{#1}}
\def\l{\ell}
\def\min{{\rm min}}
\def\max{{\rm max}}
\def\flu{{\rm fl}}
\def\c{{\rm c}}
\def\phi{\varphi}
%
%
%
\begin{center}
   {\bf \Large 
 	Spontaneous Breaking of Translational Invariance and Spatial
Condensation in Stationary States on a Ring\\[2mm]
II. The Charged System and the Two-component Burgers Equations.
}\\[15mm]
Peter F. Arndt$\mbox{}^{\star}$
and Vladimir Rittenberg$\mbox{}^{\S}$
\\[7mm]
$\mbox{}^\star$Physics Department\\
University of California at San Diego, La Jolla CA 92093-0319\\[1mm] 
Institute for Theoretical Physics\\
University of California at Santa Barbara, Santa Barbara CA 93106-4030\\[3mm]
$\mbox{}^\S$
Physikalisches Institut \\Nu{\ss}allee 12,
53115 Bonn, Germany\\[10mm]
\end{center}
\renewcommand{\thefootnote}{\arabic{footnote}}
\addtocounter{footnote}{-1}
\vspace*{2mm}
%
%
We further study the stochastic model discussed in Ref.\cite{bib2} in which
positive and negative particles diffuse in an asymmetric, CP invariant way
on a ring. The positive particles hop clockwise, the negative
counter-clockwise and oppositely-charged adjacent particles may swap
positions. We extend the analysis of this model to the case when the
densities of the charged particles are not the same.
 The mean-field equations describing the model are coupled nonlinear
differential equations that we call the two-component Burgers equations. We
find roundabout weak solutions of these equations. These solutions are  
used to describe the properties of the stationary states of the stochastic  
model.
 The values of the currents and of various two-point correlation functions
obtained from Monte-Carlo simulations are compared with the mean-field
results. Like in the case of equal densities, one finds a pure phase, a
mixed phase and a disordered phase.

\vspace{1cm}
\noindent
NSF-ITP-01-05
\newpage
\setcounter{page}{1}
\pagestyle{plain}
\newcommand{\figIII}[1]{
\begin{figure}[#1]
	\setlength{\unitlength}{1mm}
	\begin{picture}(155,90)(-20,-2)
	\put(114,-1){\makebox{$\rho_{\pm}$}}
	\put(0,77){\makebox{$j_{\pm}$}}
	\put(0,0){
        	\makebox{
			\setlength\epsfxsize{12cm}
                	\epsfbox{fig3.epsf}}}
%
	\end{picture}
	\caption{
		\label{figIII}
The current densities as a function of the densities for $q=1.2$,
$\lambda=1$, $L=1000$, $p=0.4$, $m=0.1$ 
(lattice mean-field results). Using Eq.(\ref{eq26})
one finds $V=0.11878$.
        }
\end{figure}
}
%
%
\newcommand{\figI}[1]{
\begin{figure}[#1]
	\setlength{\unitlength}{1mm}
	\begin{picture}(155,90)(-20,-2)
	\put(114,-1){\makebox{$L$}}
	\put(0,77){\makebox{$V$}}
	\put(0,0){
        	\makebox{
			\setlength\epsfxsize{12cm}
                	\epsfbox{fig1.epsf}}}
%
	\end{picture}
	\caption{
		\label{figI}
The drift velocity $V$ as a function of the size of the system $L$.
The upper curve corresponds to the mixed phase, the intermediate one to
$q=1$ and the lowest one to the pure phase.
        }
\end{figure}
}
%
%
\newcommand{\figIV}[1]{
\begin{figure}[#1]
	\setlength{\unitlength}{1mm}
	\begin{picture}(155,90)(-20,-2)
	\put(114,-1){\makebox{$L$}}
	\put(0,77){\makebox{$D$}}
	\put(0,0){
        	\makebox{
			\setlength\epsfxsize{12cm}
                	\epsfbox{fig4.epsf}}}
%
	\end{picture}
	\caption{
		\label{figIV}
The diffusion coefficient $D$ as a function of the size of the
system $L$. The upper curve corresponds to the mixed phase, the intermediate
one for $q=1$ and the lowest one to the pure phase.
        }
\end{figure}
}
%
%
\newcommand{\figVIImc}[1]{
\begin{figure}[#1]
	\setlength{\unitlength}{1mm}
	\begin{picture}(155,90)(-20,-2)
	\put(120,-1){\makebox{$L$}}
	\put(25,70){\makebox{$\av{j_+}$}}
	\put(31,18){\makebox{$3 \av{j_-}$}}
	\put(0,0){
        	\makebox{
			\setlength\epsfxsize{12cm}
                	\epsfbox{fig7mc.epsf}}}
%
	\end{picture}
	\caption{
		\label{figVIImc}
$\av{j_+}$ and $3 \av{j_-}$ from Monte-carlo data for $q=1.2,\lambda=1,p=0.4,m=0.1$.
the lines come from mean-field in the $L\rightarrow\infty$ limit.
        }
\end{figure}
}
%
%
\newcommand{\figII}[1]{
\begin{figure}[#1]
	\setlength{\unitlength}{1mm}
	\begin{picture}(155,90)(-20,-2)
	\put(114,-1){\makebox{$q$}}
	\put(0,77){\makebox{$L_{\rm min}$}}
	\put(0,0){
        	\makebox{
			\setlength\epsfxsize{12cm}
                	\epsfbox{fig2.epsf}}}
%
	\end{picture}
	\caption{
		\label{figII}
The minimal length of the system to have a condensate for different $q$
and $\lambda=1,p=m=0.2$. 
The data is very good fitted by $L=17(q_c-q)^{-3/2}$.
        }
\end{figure}
}
%
%
\newcommand{\figIX}[1]{
\begin{figure}[#1]
	\setlength{\unitlength}{1mm}
	\begin{picture}(155,90)(-20,-2)
	\put(114,-1){\makebox{$L$}}
	\put(0,77){\makebox{$V(n)$}}
	\put(0,0){
        	\makebox{
			\setlength\epsfxsize{12cm}
                	\epsfbox{fig9.epsf}}}
%
	\end{picture}
	\caption{
		\label{figIX}
The drift velocity $V(n)$ for the $n$-th condensate for $q=1.2$,
$\lambda=1$, $p=0.4$, $m=0.1$ and $n=1,2 ,\dots,16$ 
(from bottom to top). The first
condensate appears for $L>60$. The maximal drift velocity is about 0.23. 
        }
\end{figure}
}
%
%
\newcommand{\figX}[1]{
\begin{figure}[#1]
	\setlength{\unitlength}{1mm}
	\begin{picture}(155,90)(-20,-2)
	\put(114,-1){\makebox{$y$}}
	\put(0,77){\makebox{$c_{0,0},c_{+,-}$}}
	\put(0,0){
        	\makebox{
			\setlength\epsfxsize{12cm}
                	\epsfbox{fig10.epsf}}}
%
	\end{picture}
	\caption{
		\label{figX}
The correlation functions $c_{0,0}$ (top) and $c_{+,-}$ (bottom) for 
$q=1.2,\,\lambda=1,\,p=0.4,\,m=0.1$ and $L=200$.
Each curve consists of the Monte-Carlo and mean-field data lying perfectly upon each other.
        }
\end{figure}
}
%
%
\newcommand{\figXa}[1]{
%
\begin{figure}[#1]
	\setlength{\unitlength}{1mm}
	\begin{picture}(155,90)(-20,-2)
	\put(114,-1){\makebox{$y$}}
	\put(0,77){\makebox{$c_{+,0},c_{0,-}$}}
	\put(0,0){
        	\makebox{
			\setlength\epsfxsize{12cm}
                	\epsfbox{fig10a.epsf}}}
%
	\end{picture}
	\caption{
		\label{figXa}
The correlation functions $c_{+,0}$ (top) and $c_{0,-}$ (bottom) for 
$q=1.2,\,\lambda=1,\,p=0.4,\,m=0.1$ and $L=200$.
Each curve consists of the Monte-Carlo and mean-field data lying perfectly upon each other.
The drift velocity is $\lambda \lim_{y\rightarrow 0} (c_{+,0}-c_{0,-})/(1-p-m)$.
For equal densities all four curves are lying on top of each other.
        }
\end{figure}
}
%
%
%
\newcommand{\figXI}[1]{
\begin{figure}[#1]
	\setlength{\unitlength}{1mm}
	\begin{picture}(155,90)(-20,-2)
	\put(40,37){\makebox{$(j_+)$}}
	\put(40,19){\makebox{$(\rho_+)$}}
	\put(40,09){\makebox{$(j_-)$}}
	\put(50,08){\makebox{$(\rho_-)$}}
	\put(115,37){\makebox{$(j_+)$}}
	\put(115,19){\makebox{$(\rho_+)$}}
	\put(115,09){\makebox{$(j_-)$}}
%
	\put(80,72){\makebox{$(\rho_+)$}}
	\put(80,09){\makebox{$(j_+,j_-)$}}
	\put(80,23){\makebox{$(\rho_-)$}}
	\put(114,-1){\makebox{$y$}}
	\put(-15,77){\makebox{$\rho_+,\rho_-,3j_+,3j_-$}}
	\put(0,0){
        	\makebox{
			\setlength\epsfxsize{12cm}
                	\epsfbox{fig11.epsf}}}
%
	\end{picture}
	\caption{
		\label{figXI}
The profiles of the densities and currents (scaled by a factor 3)
obtained numerically on a lattice with 200 sites. The other data are like in 
Fig.\protect\ref{figIII}.
The condensate is drifting to the left.
        }
\end{figure}
}
%
%
\newcommand{\figXX}[1]{
\begin{figure}[#1]
	\setlength{\unitlength}{1mm}
	\begin{picture}(155,90)(-20,-2)
	\put(114,-1){\makebox{$q$}}
	\put(0,77){\makebox{$\av{j}$}}
	\put(0,0){
        	\makebox{
			\setlength\epsfxsize{12cm}
                	\epsfbox{fig20.epsf}}}
%
	\end{picture}
	\caption{
		\label{figXX}
The current of positive (top), negative particles (bottom) 
from Monte-Carlo Simulations 
and from mean-field calculations in the limit $L\rightarrow\infty$ 
(dashed lines) and homogeneous mean-field (solid lines).
The parameters are $\lambda=1,\,p=0.4,\,m=0.1$ and $L=800$.
        }
\end{figure}
}
%
%
%
\newcommand{\figIIb}[1]{
\begin{figure}[#1]
	\setlength{\unitlength}{1mm}
	\begin{picture}(155,90)(-20,-2)
	\put(114,-1){\makebox{$q$}}
	\put(0,77){\makebox{$a_{\rm min}$}}
	\put(0,0){
        	\makebox{
			\setlength\epsfxsize{12cm}
                	\epsfbox{fig2b.epsf}}}
%
	\end{picture}
	\caption{
		\label{figIIb}
The length of the condensate when condensation starts for different $q$
and $\lambda=1,p=m=0.2$.
        }
\end{figure}
}
%
%
\newcommand{\figF}[1]{
\begin{figure}[#1]
	\setlength{\unitlength}{1mm}
	\begin{picture}(145,135)(-20,20)
	\put(15,77){\vector(0,-1){10}}
	\put(11,72){\makebox{$t$}}
	\put(0,0){
        	\makebox{
			\setlength\epsfxsize{12cm}
                	\epsfbox{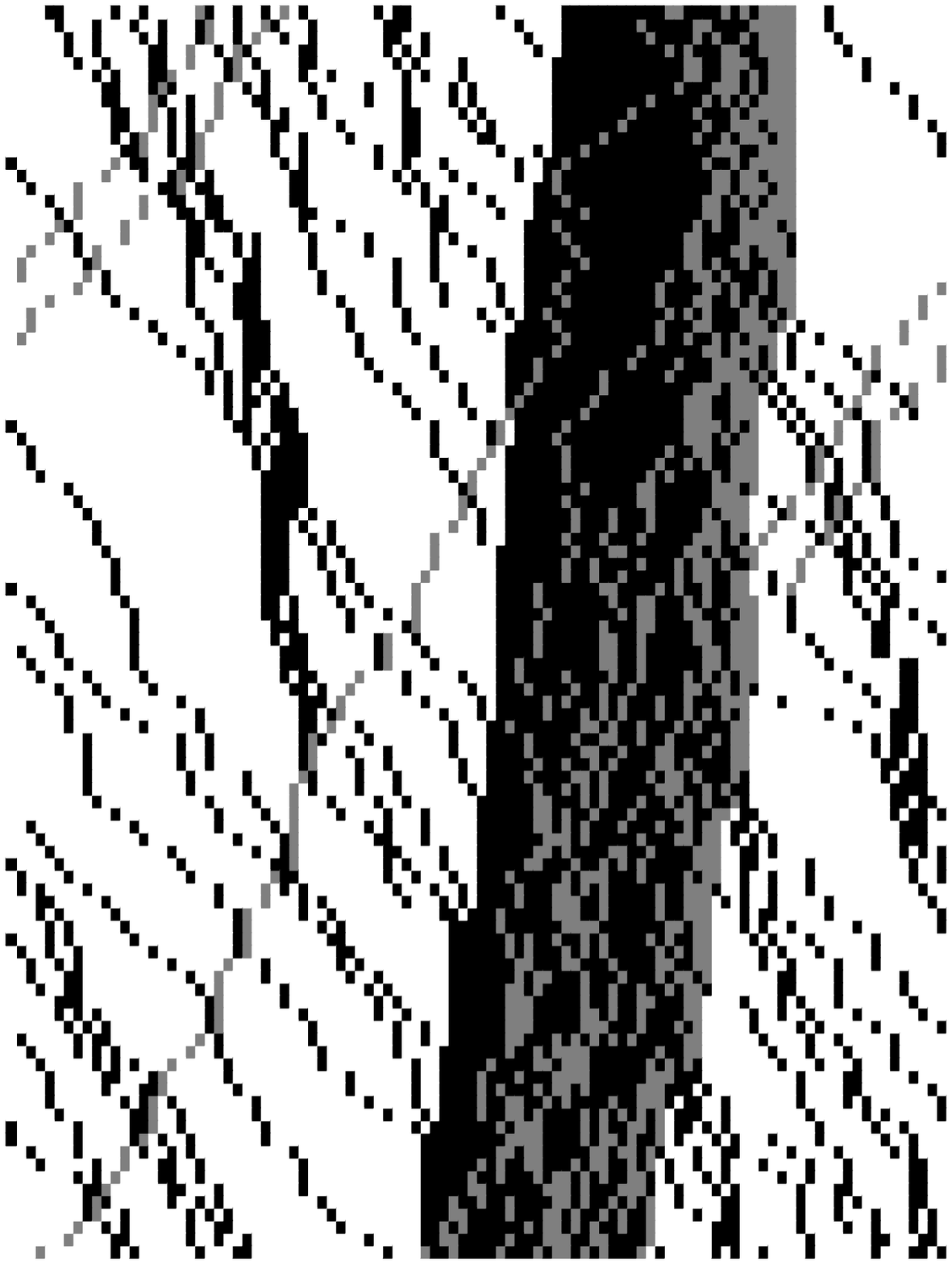}}}
%
	\end{picture}
	\caption{
		\label{figF}
A Monte-Carlo ``film'' for $q=1.2, \lambda=1, p=0.3, m=0.1$ (mixed
phase), $L=100$. The time $t$ is running from the top to the bottom of the figure.
The positive particles are depicted black, the negative ones-gray.
        }
\end{figure}
}
%
%
\newcommand{\figXV}[1]{
%
\begin{figure}[#1]
	\setlength{\unitlength}{1mm}
	\begin{picture}(155,90)(-20,-2)
	\put(114,-1){\makebox{$L$}}
	\put(0,77){\makebox{$V$}}
	\put(0,0){
        	\makebox{
			\setlength\epsfxsize{12cm}
                	\epsfbox{fig15.epsf}}}
%
	\end{picture}
	\caption{
		\label{figXV}
The ``drift velocity'' for 
$q=1.2,\, \lambda=1,\,p=0.4,\,m=0.1$ from the alegbraic approach $V=(\av{j_+}-\av{j_-})/(1-p-m)$ (points) 
and mean-field (line). 
The straight line gives the mean-field value for the limit $L\rightarrow\infty$.
        }
\end{figure}
}
%
%
\section{Introduction} \label{sec1}

Phase transitions in stationary (far away from equilibrium) states present
still many open questions. Some time ago \cite{bib1}, we have introduced a simple
one-dimensional three-state model which has interesting properties. In this
model one takes a ring with $L$ sites, the positive particles hop clockwise with
a rate $\lambda$:
	\begin{equation}
	(+)(\,0\,) \rightarrow (\,0\,)(+)   
	\label{eq1}       
	\end{equation}                    
((0) is a vacancy). The negative particles hop counter-clockwise with the
same rate $\lambda$
	\begin{equation}
	(\,0\,)(-) \rightarrow (-)(\,0\,)                                                 
	\end{equation}
and oppositely charged adjacent particles may exchange positions
	\begin{eqnarray}
	(+)(-) \rightarrow (-)(+)\qquad  \mbox{ with rate }q\\            
	(-)(+) \rightarrow (+)(-)\qquad  \mbox{ with rate }1    
	\label{eq3}                 
	\end{eqnarray}
 This model is translational invariant and the numbers of positive and
negative particles are conserved. A detailed study of this model was done
in Ref. \cite{bib2} where we have considered the case in which the densities $p$ 
(positive particles) and $m$ (negative particles) were taken equal ($p=m=
\rho$). Obviously
	\begin{equation}
	p+m+v = 1                                   
	\end{equation}
where $v$ is the density of vacancies. Using mean-field analysis, Monte-Carlo
simulations and representations of the quadratic algebra given by the
rates describing the processes (\ref{eq1}--\ref{eq3}) we have shown that the model has 
three phases. For $q<1$ (independent on $\lambda$ and the densities) translational
invariance is spontaneously broken (pure phase). In this phase, in the
thermodynamical limit, the system organizes itself into a block of positive
particles followed by a block of negative particles and one block of
vacancies. All blocks are pinned in an arbitrary position on the ring. For
$1<q<q_c(\lambda,\rho)$ (called mixed phase) one has Bose-Einstein condensation
in coordinate space (spatial condensation). In this phase one has again
charge segregation but translational invariance is not broken. To understand
better what happens (for finite but large $L$) 
it is interesting to start with the mean-field results.
Neglecting fluctuations,  the mean-field solutions, which break translational
invariance, show two pinned adjacent blocks. One block (that we called
condensate) contains only charged particles (no vacancies) with 
inhomogeneous distributions and a block (called fluid)  which contains all the
vacancies and where the particles have homogeneous distributions (see Ref. \cite{bib2}
Fig.25). As a result of fluctuations, the whole structure (condensate+fluid) makes
a Brownian motion on the ring and the stationary probability distribution
function is obtained taking with equal probability, the two adjacent blocks
anywhere on the ring. For $q>q_c$, one is in the disordered phase; there is no
charge segregation and the particles are distributed uniformly. In mean-field,
one gets:
	\begin{equation}
	q_c = 1+\frac{4\lambda\rho}{1+2\rho}
	\label{eq10}
	\end{equation}
 As shown by one of us \cite{bib3}, one can use the distributions of the zeroes of
the partition function in the grand canonical ensemble as a function of
fugacity \`a la Lee and Yang \cite{bib4}, in order to get good estimates of $q_c$ (for
other methods, see Ref. \cite{bib2}).

 The aim of this paper is to further study this model in the case of unequal
densities. As we are going to see, the physics is different. Beside our
physical motivation, we have also a mathematical one. The mean-field
equations given by the present model, are coupled nonlinear differential 
equations and as we are going to show, we are able to find exact weak \cite{bib5}
solutions of these equations. We believe that without knowing the physics of the
model it would have been very difficult to guess them. Since in the absence of
vacancies, the coupled equations reduce to the Burgers equation \cite{bib6}, we
call the mean-field equations ``The two-component Burgers equations''. In
Ref. \cite{bib2}, for equal densities, we found stationary solutions of these equations.
In the case of unequal densities we find the same three phases. Now however,
the spatial structure (which exists in the pure
and mixed phases) drifts with constant velocity around the circle. In the
large $L$ limit, the drift velocity vanishes exponentially in the pure
phase, like $1/L$ for $q=1$ and is finite for $1<q<q_c(\lambda,p,m)$.
 
 From the mean-field results one can be tempted to conclude that one has
no stationary solutions at all in the case of unequal densities even if
we take into account fluctuations. Actually one can be left with this
feeling from Ref. \cite{bib7} where a different model was considered. That
stationary solutions exist in our problem is known from the existence
\cite{bib1,bib2} of a quadratic algebra with known representations and one can obtain
in this way exact solutions for the probability distribution for the
stationary state.

 In order to understand the physics of the problem it is instructive to
have a look at the time evolution of a Monte-Carlo run which can be seen
in Fig.\ref{figF}. The parameters are chosen such that one is in the mixed phase.

\figF{bt}

 One notices that at a given time $t$ one has two regions. In one of them
(the condensate), there are no vacancies and that the boundaries are sharp.
This allows a definition of the width of the condensate: one takes the
last vacancy seen in the simulation on the left side and the first
vacancy seen on the right side. In the condensate the particles are not 
uniformly distributed: the positive ones are predominantly on the left
side and the negative ones are on the right side. The second region (the
fluid) contains all the vacancies and the charged particles are
distributed at random. If we know look at the figure downwards, one
notices that the condensate drifts to the left with a constant velocity
which can be estimated following for example the last vacancy on the
left side of the condensate.
Notice also that in the fluid, the positive particles drift to the right
whereas the negative particles drift to the left.

 In Fig.\ref{figI} we show the average values of the drift velocities obtained 
looking at the motion of the condensate
as a function of the size of the system $L$ in the pure phase ($\lambda=1,
q=0.8$), for $q=1$ ($\lambda=1$) and in the mixed phase ($\lambda=1, q=1.2$). We have
taken $p=0.4$ and $m=0.1$.  

\figI{bt}

In the pure phase, the velocity vanishes exponentially. A fit (also
shown in Fig.\ref{figI}) to the data
gives:
	\begin{equation}
	V= 0.2742\exp(-0.02615L)        	
	\end{equation}
For $q=1$, the drift velocity vanishes algebraically:
	\begin{equation}
	V= 9.602/L                 
	\label{eq18}
	\end{equation}
whereas in the mixed phase the fit gives:
	\begin{equation}
	V= 0.116 + 6.9658/L                   
	\end{equation}

\figIV{bt}


We now return to Fig.\ref{figF} and notice that the condensate does not only drift
to the left but also does a Brownian motion. If one follows again the
last vacancy on the left side of the condensate, it behaves like a random
walker and therefore one can determine the diffusion coefficient $D$.  This
was done for different sizes of the system and the results are shown in
Fig.\ref{figIV} for the same parameters as those used in Fig.\ref{figI}.
 We notice that similar
to the drift velocity $V$, for large values of $L$, the 
diffusion coefficient converges to a constant value in the mixed phase.
vanishes algebraically for $q=1$ and exponentially in the pure phase.

 As a result of these observations, it is clear that in the pure phase, in
the thermodynamical limit, one gets three adjacent pinned blocks and like in
the case of equal densities, translational invariance is broken. The situation
is different in the mixed phase. If we follow again the last vacancy on
the left side of the condensate, since it moves like a random walker with
a finite diffusion coefficient, the dispersion increases with time up to
when it reaches the perimeter of the ring. For later times one can find
this vacancy anywhere on the ring. This implies that in the stationary
state, one finds with equal probability the structure condensate+fluid in any
position on the ring. The existence of such a structure obviously implies charge
segregation (long range correlations) without breaking of translational
invariance. 

 We can understand better now what is going to be the role of mean-field
calculations. On one hand, one can estimate non-equilibrium quantities like
the drift velocity, on the other hand one can obtain the condensate+fluid
structure. Once the latter is known, one can obtain the probability distribution
of the stationary state taking as mentioned, with equal probability the charge
distribution obtained in mean-field in any position on the ring.

 The paper is organized as follows. In Sec.\ref{sec2} we consider the mean-field
equations for the case $q>1$. Those are the two-component Burgers equations.
In this section we shortly remind the reader some of the results obtained in
Ref. \cite{bib2} for the case of equal densities and also present some new ones.
In Sec.\ref{sec3} we find exact weak solutions of the two-component Burgers
equations for the case of unequal densities which tell us under which
conditions the condensation phenomenon appears. In this way one can obtain the
function $q_c(\lambda,p,m)$. Moreover one obtains the drift velocity $V$ and the
charge distributions in the condensate. In Sec.\ref{sec4} we discuss the case $q=1$,
this case is much easier to study than the $q>1$ case. We have not considered the
case $q<1$ since it can be studied along similar lines (see Ref. \cite{bib2} for the case
$p=m$). 

 In Sec.\ref{sec5} we discuss the results of Monte-Carlo simulations and compare them 
with the results obtained in mean-field used as described above. We also
give some  results using the algebraic approach \cite{bib2} which is, in the
case of unequal densities, harder to apply.
The conclusion is presented in Sec.\ref{sec6}.

 Since this work is the continuation of the one presented in Ref. \cite{bib2}, this
paper is not self-contained and it can't be understood without first reading
Ref. \cite{bib2} unless one is interested only in the solutions of the two-component
Burgers equations described in Secs.\ref{sec2}, \ref{sec3} and \ref{sec4}.
 
  We would like to mention that the model described above was subsequently
studied again on a ring, in the case in which only one vacancy (seen as a
defect) is present \cite{bib7} and for a special case of open boundaries \cite{bib8}.

\section{The two-component Burgers equations ($q>1$)} \label{sec2}
The mean-field equations in the continuum of the model defined by
Eqs.(\ref{eq1})--(\ref{eq3}) are (Ref. \cite{bib2} Appendix B):
	\begin{equation}
	\frac{\partial j_\pm}{\partial x} \pm \frac{\partial \rho_\pm}{\partial t}=0
	\label{eq4}
	\end{equation}
where $\rho_\pm$ are the local densities of the charged particles ($\rho_0$ is
the local density of the vacancies) satisfying the obvious relation:
	\begin{equation}
	\rho_++\rho_-+\rho_0=1
	\end{equation}
$j_+$ is the local current density of positive particles, $j_-$ is minus the 
local current density of negative particles (the minus sign is convenient if one looks 
at stationary states in the case of equal densities since with this choice
of signs one gets $j_+=j_-$). The expressions of the current densities
are:
	\begin{equation}
	\fl
	j_\pm=
	\lambda \rho_\pm \rho_0 +(q-1) \rho_+ \rho_-
	+\frac1{2}\left[ 
	(q+1) \left( \rho_+ \frac{\partial\rho_-}{\partial x} -\rho_- \frac{\partial\rho_+}{\partial x} \right)
	\pm \lambda \left( \rho_\pm \frac{\partial\rho_0}{\partial x} -\rho_0 \frac{\partial\rho_\pm}{\partial x} \right)
	\right]
	\label{eq7}
	\end{equation}
We are looking at periodic solutions of the Eqs. (\ref{eq4}):
	\begin{equation}
	\rho_\pm(x,t)=\rho_\pm(x+L,t)
	\end{equation}
where $L$ is the perimeter of the ring. We will also be interested in
solutions with a drift velocity $V$ counter clockwise (see Fig.\ref{figF}):
	\begin{equation}
	\rho_\pm(x,t)=\rho_\pm(x+V t)
	\label{eq25}
	\end{equation}
which implies that in the moving frame, the quantities
	\begin{equation}
	C_\pm=\pm V \rho_\pm + j_\pm
	\label{eq26}
	\end{equation}
are independent of $x$. It is useful to define
	\begin{equation}
	y=\frac{x}{L}
	\qquad\qquad 0\leq y\leq 1
	\label{eq27}
	\end{equation}
and 
	\begin{equation}
	\fl
	J_\pm =\frac{C_\pm}{q-1}
	\,,\,
	W=\frac{V}{q-1}
	\,,\,
	\l_\pm=\frac{j_\pm}{q-1}
	\,,\,
	\nu=\frac{q+1}{2L(q-1)}
	\,,\,
	\eta=\frac{\lambda}{q-1}
	\,,\,
	\mu=\frac{\eta}{2L}
	\label{eq5}
	\end{equation}
In Eq.(\ref{eq5}) we have assumed that $q>1$. Rescaling the time, instead of
Eqs. (\ref{eq4}),(\ref{eq7})--(\ref{eq26}) one finds:
	\begin{equation}
	\frac{\partial \l_\pm}{\partial y} \pm \frac{\partial \rho_\pm}{\partial t}=0
	\label{eq29}
	\end{equation}
	\begin{equation}
	\fl
	\l_\pm=
	\eta \rho_\pm \rho_0 +\omega  \rho_+ \rho_-
	+\nu 
	\left( \rho_+ \frac{\partial\rho_-}{\partial y} -\rho_- \frac{\partial\rho_+}{\partial y} \right)
	\pm \mu \left( \rho_\pm \frac{\partial\rho_0}{\partial y} -\rho_0 \frac{\partial\rho_\pm}{\partial y} \right)
	\label{eq210}
	\end{equation}
	\begin{equation}
	\l_\pm(y,t)=\l_\pm(y+1,t)
	\label{eq211}
	\end{equation}
	\begin{equation}
	\l_\pm(y,t)=\l_\pm(y+Wt)
	\end{equation}
	\begin{equation}
	J_\pm=\pm W \rho_\pm \pm \l_\pm
	\label{eq213}
	\end{equation}
where $\omega=1$.
Since the numbers of positive and negative particles are conserved, we are
looking for solutions for given densities $p$ and $m$:
	\begin{equation}
	\int_0^1 \rho_+ \,\d y =p
	\,,\qquad
	\int_0^1 \rho_- \,\d y =m
	\label{eq214}
	\end{equation}
We call the Eqs. (\ref{eq210})--(\ref{eq211}) the two-component Burgers equations. We
have coined this name to the equations since if the density of vacancies $\rho_0$
is equal to zero, we can make in Eqs.(\ref{eq210}) the substitution
	\begin{equation}
	\l_\pm=\frac12 \pm f
	\label{eq215}
	\end{equation}
and see that $f$ satisfies the Burgers equation \cite{bib6}
	\begin{equation}
	\frac{\partial f}{\partial t}=
	\nu \frac{\partial^2 f}{\partial y^2}
	+\zeta  f \frac{\partial f}{\partial y}
	\label{eq216}
	\end{equation}
where $\zeta=2$ and $\nu$ is the damping constant whose physical interpretation in our
model can be read of from Eq.(\ref{eq5}). One notices that in the infinite volume
limit, the damping constant vanishes and one obtains the inviscid Burgers
equation \cite{bib6}. The constant $\mu$ appearing in Eqs.(\ref{eq210}) will be called
moisture constant. Notice (see Eq.(\ref{eq5})) that the 
moisture constant also vanishes in the large $L$ limit.

An important result of our investigation is that one is able to give
exact solutions to the two-component Burgers equation. Let us review the
results obtained in Ref. \cite{bib2} (Section 6). For equal densities, in the mixed
phase, inspired by the results obtained numerically (see Fig.25), one can
obtain the stationary solutions in the following way. We divide the
segment $0<y<1$ into two smaller segments: $0<y<a$ where we have a condensate
(no vacancies) with a non-uniform distributions of the charged particles
and a second segment of length $b$, ($a+b = 1$) where the particles have a
uniform distribution (fluid). Therefore in the condensate one can use the
Burgers equation whereas in the fluid one has almost nothing to compute.
Numerical investigation teach us how to sew the two segments: the
density of positive particles is smooth at the right-side of the condensate
($y=a$) but not at the left-side ($y=0$) where it is discontinuous. For the
negative particles the discontinuity is at $y=a$ as a consequence of the CP
invariance of the problem. 
The solutions obtained in this way are weak solutions of the differential
equations which correspond to taking the moisture constant $\mu$ equal to 
zero. They represent an excellent approximation not only of the 
concentration profiles obtained numerically for the mean-field equations
but also of the concentration profiles seen in computer simulations
(see Fig.8 and 9 in Ref. \cite{bib2}).

The values of $a$ are dependent on $\nu$, $\eta$ and
the density $p=m=\rho$. The domain in the space of the variables $\nu$, 
$\eta$ and $\rho$ where solutions exist define the mixed phase. In the 
thermodynamic limit ($\nu=0$) one obtains for $b$ the value 
	\begin{equation}
	B=\frac{v(1+2\eta)}{2(1+\eta)}
	\label{eq11}
	\end{equation}
The result mentioned in section \ref{sec1} for $q_c$ (see Eq.(\ref{eq10})) is obtained from
Eq.(\ref{eq11}) when takes $B=1$ (the condensates disappears). To the results
obtained in Ref. \cite{bib2} we would like to add some supplementary data which are
relevant also for the case of unequal densities. For a given value of $\eta$
there is a maximum value of $\nu$ ($\nu_{\max}$) beyond which the condensation does
not take place anymore. Moreover, one would expect that for $\nu_{\max}$ the
length of the condensate $a$ is zero and that $a$ increases as the
damping constant decreases. This is not the case: for $\nu=\nu_{\max}$ one gets
a non-vanishing value for $a$ equal to $a_\min$.

\figII{bt}

\figIIb{bt}

In Figs.\ref{figII} and \ref{figIIb} one can see this phenomenon. We work with
the variables $q$, $\lambda$ and $L$ instead of $\nu$ and $\eta$ (three variables
instead of two) not only because our physical intuition is in these
variables but also because in section \ref{sec5} we will look into the stochastic
problem where, unlike in mean-field with $\mu=0$,  on has all the three variables.

 One notices that as we approach the critical point ($q_c=1.5714$ as obtained
from Eq.(\ref{eq10})), the minimal length $L_\min$ diverges and the the smallest
condensate $a_\min$ vanishes.


 
 
We should stress again that
the solutions described above are weak solutions for the stationary states
of the two-component Burgers equation. The current is constant on the
segment $0<y<1$ but the densities are discontinuous. One can ask the question
how do these discontinuous solutions look like if one looks at the
``microscopic'' problem on the lattice both in mean field and in Monte-Carlo
simulations. As one can see in Ref. \cite{bib2}, Fig.25 or in the present paper Figs.
\ref{figF} and \ref{figXI}, there are two dramatic changes in the densities over not more than 
a few
sites for any lattice size $L$. If one works with the $y=x/L$ variable,
these localized, abrupt, changes lead to discontinuous solutions.

 One can ask if in the mixed phase the condensate+fluid solutions are the
single ones. The answer is no. Obviously one has a solution where all the
densities are constant (pure fluid). More interestingly, for a given value
of $\nu$, $\eta$ and $\rho$ one can have solutions with $n$ condensates
(fluid+condensate+$\cdots$+fluid+condensate) all of them with the same
distribution of charged particles and located anywhere on the ring. It is
easy to show that in order to get the solutions with $n$ condensates one has
to use the solutions obtained  for a single condensate in which
we take the damping constant equal to $n\nu$. We didn't study the
problem of the stability of these solutions. What we do know is that we have also
obtained numerically these solutions by using the
time-dependent mean field equations (see \cite{bib2} Eq.(B2)) and that we have
obtained, depending on the initial conditions, solutions with no condensate,
with one condensate and with two condensates. We didn't take large
lattices where one could have seen more than two condensates.  

One comment is in place concerning the multi-condensate solutions of mean-field.
In the presence of fluctuations, each condensate behaves like
a random walker who sooner or later will meet another random walker.
When two condensates meet, they melt into a single condensate so that 
if one starts with $n$ condensates, we end up with a single one. This
is seen in Monte-Carlo simulations.

\section{Roundabout solutions of the two-component Burgers equations $(q>1)$} \label{sec3}
Before describing roundabout solutions of the differential equations (\ref{eq29})
obtained in the case of unequal densities for the positive and negative
particles, it is interesting to give some results obtained looking at
large time solutions of the mean-field equations on the lattice (Ref. \cite{bib2}
Eq.(B2)) obtained numerically. In Fig.6 we plot the current densities as a
function of the densities of the positive and negative particles for a
lattice with 1000 sites for $q=1.2$, $\lambda=1$ and $p=0.4$ and $m=0.1$. We are
going to see later that we are in the mixed phase. We notice that in
agreement with Eq.(\ref{eq27}) one obtains straight lines and from their slopes
one can with high precision the drift velocity $V$.

\figIII{bt}

\figXI{bt}
 
In Fig.\ref{figXI} we show the
distributions of the current densities and of the particle densities for a
shorter lattice ($L=200$) keeping all the other data unchanged. Like in the
case of equal densities, one observes a condensate (a region without
vacancies) and a fluid with uniformly distributed charged particles.
 Based on these numerical observations one can start to look for
roundabout solutions of Eqs.(\ref{eq29}), taking into account Eqs. (\ref{eq211})--(\ref{eq214}).
We first start with the condensate ($0<y<a$) where we have no vacancies.
Denoting 
	\begin{equation}
	\rho_\pm^\c=\frac{1+W}{2}\pm u
	\label{eq31}
	\end{equation}
	\begin{equation}
	J=\frac{J_++J_-}{2}
	\end{equation}
from Eq.(\ref{eq213}) we obtain:
	\begin{equation}
	J_+-J_-=W
	\label{eq33}
	\end{equation}
and from Eqs.(\ref{eq213}) and (\ref{eq210}) we get:
	\begin{equation}
	\alpha^2=4J-1-W^2=-4(u^2+\nu \frac{\partial u}{\partial y})
	\label{eq34}
	\end{equation}
We have denoted by $\rho_\pm^\c$  the densities of the charged particles in the
condensate. Obviously
	\begin{equation}
	\rho_+^\c+\rho_-^\c=1
	\end{equation}
We can integrate Eq.(\ref{eq34}) and find:
	\begin{equation}
	u=\beta \nu \tan [\beta(a \phi -y)]
	\end{equation}
where $\phi$ is an arbitrary constant and
	\begin{equation}
	\beta=\frac{\alpha}{2\nu}
	\end{equation}

At this point $J$, $W$ and $\phi$ are unknown constants which still have to be
fixed. We now consider the fluid ($a<y<1$) where all the vacancies are
concentrated. In this domain the densities $\rho^\flu_\pm$ and $\rho^\flu_0$ are
constant (independent of $y$) and verify the relation:
	\begin{equation}
	\rho_+^\flu+\rho_-^\flu+\rho_0^\flu=1
	\end{equation}
We obviously have
	\begin{equation}
	\rho_0^\flu=v/b
	\label{eq39}
	\end{equation}
where $v$ is the density of vacancies and $b$ the length of the fluid. From 
Eq.(\ref{eq210}) we get:
	\begin{equation}
	\l^\flu_\pm=\eta \rho^\flu_\pm \rho^\flu_0 + \rho^\flu_+\rho^\flu_-
	\end{equation}
 We denote
	\begin{equation}
	\Delta=\rho^\flu_+ - \rho^\flu_-
	\end{equation}
and get
	\begin{equation}
	W=\eta \Delta \rho^\flu_0
	\end{equation}
and
	\begin{equation}
	\alpha^2=-\Delta^2(\eta-1)^2 - (\rho^\flu_0)^2(2\eta -1)+2 \rho^\flu_0 (\eta-1)
	\label{eq313}
	\end{equation}
We now sew the condensate with the fluid asking for the following
conditions to be fulfilled:
	\begin{equation}
	\rho^\c_+(a)=\rho^\flu_+
	\,,\quad
	\rho^\c_-(0)=\rho^\flu_-
	\label{eq314}
	\end{equation}

Using Eqs.(\ref{eq31}) and (\ref{eq314}) we get:
	\begin{equation}
	u(0)=\frac12-\frac{W}{2}-\rho^\flu_-
	\,,\quad
	-u(a)=\frac12+\frac{W}{2}-\rho^\flu_+
	\end{equation}
and therefore:
	\begin{equation}
	\rho^\flu_0=\beta \nu \big(
	\tan[\beta a(1-\phi)]+\tan[\beta a \phi]
	\big)
	\label{eq316}
	\end{equation}
and
	\begin{equation}
	W-\Delta =\beta \nu \big(
	\tan[\beta a(1-\phi)]-\tan[\beta a \phi]
	\big)
	\label{eq317}
	\end{equation}
We now use Eq.(\ref{eq214}) and get for $n$ condensates:
	\begin{equation}
	n \int_0^a \rho^\c_+ \d y +\rho^\flu_+ b=p
	\,,\quad
	n \int_0^a \rho^\c_- \d y +\rho^\flu_- b=m
	\label{eq318}
	\end{equation}
It is useful to denote
	\begin{equation}
	d=p-m
	\end{equation}
and from Eq.(\ref{eq318}) we obtain:
	\begin{equation}
	d=(Wa+2 \int_0^a u\, \d y)n +\Delta b
	\label{eq320}
	\end{equation}
 We have to make again a change of notations:
	\begin{equation}
	\beta a \phi =\frac\pi2-\psi_1
	\,,\quad
	\beta a (1-\phi) =\frac\pi2-\psi_2
	\end{equation}
 In the new notations Eqs. (\ref{eq316}), (\ref{eq317}), (\ref{eq320}) and (\ref{eq313}) become:
	\begin{equation}
	\rho^\flu_0=(\pi-(\psi_1+\psi_2))\frac\nu{a} [\cot \psi_2+\cot \psi_1]
	\label{eq322}
	\end{equation}
	\begin{equation}
	\Delta(\eta-1)=(\pi-(\psi_1+\psi_2))\frac\nu{a} [\cot \psi_2-\cot \psi_1]
	\label{eq323}
	\end{equation}
	\begin{equation}
	d-\Delta(\eta-(\eta-1)b)=2\nu n \, \log\left(\frac{\sin\psi_2}{\sin\psi_1}\right)
	\end{equation}
	\begin{equation}
	\frac{4\nu^2}{a^2}(\pi-(\psi_1+\psi_2))^2=-\Delta^2(\eta-1)^2- (\rho^\flu_0)^2(2\eta-1)
	+2 \rho^\flu_0 (\eta-1)
	\label{eq325}
	\end{equation}
Taking into account Eq.(\ref{eq39}), the four unknown $b$, $\psi_1$, $\psi_2$ and $\Delta$ can
be obtained from the four equations (\ref{eq322}-\ref{eq325}) once $p$ and $m$ are given.
 In this way one can determine the profiles of densities in the condensate
and the fluid, the currents and the drift velocities once we give $\eta$, $\nu$,
$p$ and $m$ . We have not determined the domain of these variables where the
solutions exist. Having in mind the stochastic process where we have used 
the mean-field solutions in order to explain the properties of the
stationary distributions, we have just looked at the values $q=1.2$,
$\lambda=1$, $p=0,4$ and $m=0.1$ in the mixed phase. As mentioned already before,
and as can be seen from the equations given above,
if we know the solutions for one condensate for any value  $L$, one 
can obtain the solutions of the problem for $n$ condensates for a lattice size $L$ taking
the solution for one condensate for a lattice of size $L/n$.  

 In Fig.\ref{figIX} we show the drift velocities for the condensates which appear 
up to $L = 1000$. One notices first that the first condensate appears for
$L=60$. This implies that that one obtains $n$ condensates when (roughly)
$L>60n$. Next one can see that for a given value of $L$ many condensates move
faster than fewer ones.

\figIX{bt}

Various applications to the stochastic process of the distributions of the charged particles
obtained
solving the transcendental equations given above can be found in Sec.\ref{sec5}.
(see Figures \ref{figX} and \ref{figXa}).

 The calculations simplify in the large $L$ limit when the friction constant
$\nu$ constant is small. Denoting the limiting values by
	\begin{equation}
	b\rightarrow B
	\,,\quad
	\Delta \rightarrow D
	\,,\quad
	\nu \rightarrow 0
	\end{equation}
and
	\begin{equation}
	C=(\eta-1)B
	\end{equation}
we obtain
	\begin{equation}
	\psi_1=\frac{2\pi\nu (\eta-1)}{1-B}
	\left[\frac{v}{C}-D\right]^{-1}
	\,,\quad
	\psi_2=\frac{2\pi\nu (\eta-1)}{1-B}
	\left[\frac{v}{C}+D\right]^{-1}
	\end{equation}
and 
	\begin{equation}
	D=\frac{d}{\eta-C}
	\,,\quad
	\rho^\flu_0=\frac{v}{B}
	\label{eq329}
	\end{equation}
 $C$ can be determined from the cubic equation
	\begin{equation}
	\fl
	C^3
	-C^2\left[2 \eta +\frac{v}{2}(\eta-1)+\frac{d^2}{2v}\right]
	+C \eta \left[\eta+v(2\eta-1)\right]
	-\frac{\eta^2 v}{2}(2\eta -1)
	=0
	\label{eq330}
	\end{equation}
 We have used the equations (\ref{eq329}) and (\ref{eq330}) in the case $q=1.2$, $\lambda=1$,
$p=0.4$ and $m=0.1$ and have obtained
	\begin{eqnarray}
	V=0.11180
	\,,\,\,
	B= 0.57929
	\,,\,\,
	D=0.11181
	\nonumber\\
	\rho^\flu_0=0.86313  
	\nonumber\\
	j_+^\flu=0.1076
	\,,\,\,
	j_-^\flu=0.0111
	\,,\,\,
	j_+^\c=j_-^\c=0.03437
	\label{eq331}
	\end{eqnarray}
Notice that the value obtained this way for $V$ is closed to the
value determined numerically on the lattice as
given in the caption of Fig.\ref{figIII}. As we are going to explain in
Section \ref{sec5}, the average values of the current densities over the ring are
very useful. In the large $L$ limit they are given by 
	\begin{equation}
	\langle j_\pm \rangle =
	(1-B) j_\pm^\c + B j_\pm^\flu
	\label{eq332}
	\end{equation}
Using the values given in (\ref{eq331}), one obtains:
	\begin{equation}
	\langle j_+ \rangle =0.07678
	\,,\,\,
	\langle j_- \rangle =0.0288  	
	\label{eq333}
	\end{equation}
These values are going to be used in Section \ref{sec5}.

\section{Mean-field solutions for $q=1$} \label{sec4}
The starting point is again the Eqs.(\ref{eq25}) and (\ref{eq26}). Making the notations 
	\begin{equation}
	\nu=\frac1{L}
	\,,\quad
	\mu=\frac1{2\lambda L}
	\,,\quad
	\l_\pm=\frac{j_\pm}{\lambda}
	\,,\quad
	W=\frac{V}{\lambda}
	\,,\quad
	J_\pm=\frac{C_\pm}{\lambda}
	\end{equation}
we find the expression (\ref{eq210}) with $\eta=1$ and $\omega=0$. 
 We repeat the procedure used in the last section assuming again that the
moisture constant $\mu$ is zero. This assumption is again based on numerical
solutions of the mean-field equations as can be seen for example in Fig.9 
of Ref. \cite{bib2} in the case of equal densities. In the region without  vacancies,
using again the substitution (\ref{eq215}) we obtain the Burgers equation (\ref{eq216})
with $\zeta=0$.

For the condensate we use again Eqs.(\ref{eq31})--(\ref{eq33}) and get:
	\begin{equation}
	J=\frac{J_++J_-}{2}
	=-\frac{W}{2}+V\rho_+^c-\nu \frac{\partial \rho_+^c}{\partial y}
	\label{eq42}
	\end{equation}
From Eq.(\ref{eq42}) we obtain the distribution of the positive particles in the
condensate ($0<y<a$)
	\begin{equation}
	\rho_+^c=A \e{\frac{Wy}{\nu}}+J_+
	\end{equation}
where 
	\begin{equation}
	A=\frac{-\rho_+(0)+\rho_+(a)}{\e{\frac{Wa}{\nu}}-1}
	\end{equation}
and 
	\begin{equation}
	J_+=
	W\left[ \frac{ \e{\frac{Wa}{\nu}} \rho_+(0)- \rho_+(a)}{\e{\frac{Wa}{\nu}}-1} \right]
	\end{equation}
 We also have:
	\begin{equation}
	J=W \left[ \frac {\e{\frac{Wa}{\nu}} \rho_+^\c(0)- \rho_+^\c(a)}
			{\e{\frac{Wa}{\nu}}-1} -\frac12 \right]
	\end{equation}
 In the fluid ($a<y<1$) where the densities are constant, we obtain
	\begin{equation}
	J_+-J_-=W(1-\rho_0^\flu) +(\rho_+^\flu-\rho_-^\flu)\rho_0^\flu
	\end{equation}
and
	\begin{equation}
	J=\frac{W}{2}(\rho_+^\flu-\rho_-^\flu)+\frac12 (1-\rho_0^\flu) \rho_0^\flu
	\label{eq48}
	\end{equation}
as well as:
	\begin{equation}
	\rho_+^\flu-\rho_-^\flu =W
	\end{equation}
 Since $J$ has to be constant over all the segment $0<y<1$, we get from
Eqs.(\ref{eq42}) and (\ref{eq48}):
	\begin{equation}
	W^2+ (1-\rho_0^\flu) \rho_0^\flu=-W+2W \rho_+^\c-2\nu \frac{\partial \rho_+^c}{\partial y}
	\end{equation}
 We now ask that the density of positive particles is smooth at the right
hand side of the condensate ($y=a$) and that the density of negative
particles is smooth at the left hand side of the condensate ($y=0$) and get
	\begin{equation}
	\rho_+^c(a)=\frac12(1+W-\rho_0^\flu)
	\,,\quad
	\rho_+^c(0)=\frac12(1+W+\rho_0^\flu)
	\end{equation}
as well as:
	\begin{equation}
	J=\frac{W^2}{2} +\frac{W}{2} \frac{\e{\frac{Wa}{\nu}}+1}{\e{\frac{Wa}{\nu}}-1}\rho_0^\flu
	\end{equation}
	\begin{equation}
	1-\rho_0^\flu=W	\frac{\e{\frac{Wa}{\nu}}+1}{\e{\frac{Wa}{\nu}}-1}
	\label{eq413}
	\end{equation}
and
	\begin{equation}
	\rho_+^\c(y)=\frac12(1+W) +  \frac{1+\e{\frac{Wa}{\nu}}-2 \e{\frac{Wy}{\nu}}}
		{2(\e{\frac{Wa}{\nu}}-1)}\rho_0^\flu
	\label{eq414}
	\end{equation}
We now use Eq.(\ref{eq318}) which fixes the densities of the charged particles
and use Eq.(\ref{eq414}) to obtain:
	\begin{equation}
	\rho_0^\flu(1-a)=v
	\label{eq415}
	\end{equation}
	\begin{equation}
	p-m=W+\rho_0^\flu \left[ a \frac{\e{\frac{Wa}{\nu}}+1}{\e{\frac{Wa}{\nu}}-1} -\frac{2\nu}{W} \right]
	\label{eq416}
	\end{equation}
($v$ is the density of vacancies). It is convenient to denote 
	\begin{equation}
	\frac{Wa}{\nu}=x
	\end{equation}
 From Eqs.(\ref{eq413}) and (\ref{eq416}) we obtain two equations for the unknown $x$ and
$a$:
	\begin{equation}
	p-m=\frac{x\nu}{a} +\frac{v a}{1-a} 
	\left[ \frac{\e{x}+1}{\e{x}-1} -\frac{2}{x} \right]
	\label{eq418}
	\end{equation}
	\begin{equation}
	1-\frac{v}{1-a}=\frac{x\nu}{a} \frac{\e{x}+1}{\e{x}-1}
	\end{equation}
(we have taken into account the relation (\ref{eq415})).

In the large $L$ limit (very small $\nu$) we obtain
	\begin{equation}
	\rho_0^\flu=1
	\end{equation}
which means that there are no charged particles in the fluid therefore the
length of the fluid is
	\begin{equation}
	B=v
	\end{equation}
From Eq.(\ref{eq418}) we get
	\begin{equation}
	\frac{p-m}{p+m}= \frac{\e{x}+1}{\e{x}-1} -\frac{2}{x}
	\end{equation}
which gives $x$ once the densities $p$ and $m$ are given. Knowing $x$ we get $W$:
	\begin{equation}
	W=\frac{x\nu}{p+m}
	\end{equation}
 It is an interesting exercise to derive the drift velocity $V$ from these
equations. For $p=0.4$ and $m=0.1$ we get $V=9.602/L$ in excellent agreement with the
the estimate obtained from Monte-Carlo simulations (\ref{eq18}). 
 One can easily derive the average (over the ring) values of the currents:
	\begin{equation}
	\langle j_+ \rangle =
	\frac{x \nu}{p+m} \left( \frac{\e{x}}{\e{x}-1}-p \right)
	\,,\,\,
	\langle j_- \rangle =	
	\frac{x \nu}{p+m} \left( \frac{1}{\e{x}-1}+m \right)
	\label{eq424}
	\end{equation}
and notice that they decrease like the inverse of the length $L$ of the
chain.

\section{ 
The mixed phase. Comparison between mean-field results and Monte-Carlo
simulations
}\label{sec5}
In the last two sections we have shown, the mean-field
equations related to the stochastic process defined by Eqs.(\ref{eq1})--(\ref{eq3}). In
Ref. \cite{bib2}  we have discussed in detail, in the case of equal densities, how to
use mean-field in order to explain the properties of the stationary states.
One assumes that in the mixed phase, one can take with equal probability
in any position of the ring the structure (condensate+fluid) obtained in
mean-field. In this section we extend this discussion for the case of
unequal densities. We limit ourselves to the case $p=0.4$, $m=0.1$, and
$\lambda=1$ only.

\figVIImc{bt}

\figXX{bt}
 In Fig.\ref{figVIImc} we give the average values 
of the currents $\langle j_+\rangle$ and $\langle j_-\rangle$, for
$q=1.2$ and various lattice sizes. Within the errors they look like they
converge to the large $L$ limit obtained in mean-filed. Since in mean-field
the current-densities are not constant (see Fig.\ref{figXI}), one has to compute the
average value of the current densities $j_+$ and $j_-$ on the ring. For the
case $q=1$ the expressions of these averages are given in Eq.(\ref{eq424}). For
$q>1$, in the infinite $L$ limit only, the calculation of the averages is
again simple since in the condensate the positive and negative particles
have a uniform distribution like in the fluid (see also Eqs.(\ref{eq332})--(\ref{eq333})). 
The calculation is more
complicated in the case of large but finite values of $L$.

 The average values of the currents for various values of $q$ are shown in
Fig.\ref{figXX}. The mean-field values are obtained as described above, the
Monte-Carlo values are obtained using results from finite lattices and
extrapolating to $L$ infinity. Similar to the case of equal densities, the
currents are very well given by mean-field up to a value $q_c$ (around 1.6).
This is the mixed phase. For $q>q_c$ the currents have a different behaviour (one
is in the disordered phase).
 
\figX{bt}

\figXa{bt}
 In Figs.\ref{figX} and \ref{figXa} we show the two-point correlation functions as
a functions of the distance $R=yL$ obtained in Monte-Carlo simulations
($L=200$) and in mean-field. In the case of correlation functions, in order
to obtain the mean-field results one had to use the results of Sec.\ref{sec3} for
$L=200$. The agreement between the simulations data and the mean-field results
is remarkable.

\figXV{bt}
 Since one is in a stationary state, one can't talk about drift velocities
which are seen in non-equilibrium or in the roundabout solutions of the
mean-field equations. Let us observe however that when we use the
mean-field results in order to derive the correlation functions given in
Fig.\ref{figXa}, for small values of $y$, one obtains the following value of the
drift velocity $V$
	\begin{equation}
	V=\lambda \lim_{y\rightarrow 0} 
	\frac{c_{+,0}-c_{0,-}}{1-p-m}
	\label{eq51}
	\end{equation}
This expression is so simple because in this limit, only the fluid and not
the condensate plays a role. Using the results of the Monte-Carlo
simulations and Eq.(\ref{eq51}) one can estimate the drift velocity. One gets $V=0.14$
which is perfectly consistent with the value obtained looking at the random
movement of the condensate (see Fig.\ref{figI} and Fig.\ref{figXV}).
 
 Up to now we have discussed only the Monte-Carlo data. In Ref. \cite{bib2} we have
also shown how to use the algebraic approach in order to obtain the
current using the grand canonical ensemble (for correlation functions this
approach gives wrong results). In the case of unequal densities one has to
deal with two chemical potentials and the calculations become difficult.
In Fig.\ref{figXV} we give the values for the quantity
	\begin{equation}
	V=
	\frac{\langle j_+\rangle-\langle j_-\rangle}{1-p-m}
	\label{eq52}
	\end{equation}
obtained using the algebraic approach and compare it to the corresponding
values obtained in mean-field. In mean-field $V$ as defined by Eq.(\ref{eq52}) is
indeed the drift velocity (use Eq.(\ref{eq26}) and compute $C_\pm$ in the fluid).
The agreement between the two calculations looks reasonable.

\section{Conclusion}\label{sec6}
In this sequel of Ref. \cite{bib2} we have introduced what we call the
two-component Burgers equations defined by Eqs.(\ref{eq29}) and (\ref{eq210}). These
equations correspond to the mean-field approximation of the stochastic
model defined in Sec.\ref{sec1} We have found roundabout weak solutions of these
equations for $\omega=1$ in Sec.\ref{sec3} and for $\eta=1$ and $\omega=0$ 
in Sec.\ref{sec4}. These
two cases correspond to the mixed phase, respectively to the $q=1$ case, of
the stochastic model. The weak solutions of the two-component Burgers equations
are obtained if we take the moisture constant $\mu=0$ in these equations.
When compared with the solutions of the equations when the moisture
constant is not zero, which can be obtained numerically, one sees that one
gets a very good approximation.

 The analytical roundabout solutions presented in Sec.\ref{sec3} allow to
determine the domain in the parameter space where the mixed phase exists.
Moreover, using these solutions one can obtain predictions for the stochastic
problem as explained in Sec.\ref{sec5}. These predictions are in good agreement
with the results of Monte-Carlo simulations on large lattices, also for the case
in which the densities of positive and negative particles are not equal. Since
Monte-Carlo simulations are much time-consuming, in the present paper we have
performed a less careful analysis of the data as compared with the case of
equal densities.   

 Before finishing this paper we will like to comment on a paper \cite{bib10}
where, in the case of equal densities, one questions the existence of the mixed
phase. If one uses the grand canonical ensemble and the algebra presented in
Ref. \cite{bib2} and two hypotheses one argues that for "cosmological" large
lattices the mixed phase disappear. For example, if one takes $\lambda=1$ and
$q=1.11$, the dimension of the lattice should be of the order $10^{70}$ in
order to "loose" the mixed phase. If the value of $q$ gets closer to 1 one get
lattice lengths of the order $10^{490}$ or more. If the calculations of Ref.
\cite{bib10} describe indeed the reality this would be a fascinating
phenomenon, certainly more interesting that the model itself. In Ref.
\cite{bib10} one has made one mathematical hypothesis which was proven correct
\cite{bib11} and a technical hypothesis according to which one neglects in the
calculation configurations with no vacancies. This second hypothesis, which was
used already in Ref. \cite{bib2}, is probably also irrelevant \cite{bib12}. Our
own doubt about the results of Ref. \cite{bib10} come from the use of the grand
canonical ensemble. As stressed in Ref. \cite{bib2}, for stochastic processes
as defined for example in Sec.\ref{sec1}, there is no grand canonical ensemble.
One can make an ad hoc definition, hope to be lucky and get the right results.
As shown in Ref. \cite{bib2}, for the two-point correlation functions in the
mixed phase, the canonical and grand canonical ensembles give very different
results (see Fig.21). It looked however that for the one-point function (the
current) the results were correct. We don't want here to defend here the
existence of the mixed phase (if it survives up to lattice lengths of the order
of $10^{70}$, it is enough for physical purposes), on the contrary, we think
that the real challenge is to clarify the problem. One more argument in the
favour of the existence of the mixed phase can be found in Ref. \cite{bib13}
where one considers the open system (no grand canonical calculation in this
case) for the same bulk parameters as those taken on the ring when one has the
mixed phase.  The input and output rates are chosen symmetric for the positive
and negative particles. One sees a first-order phase transition between a phase
where the density corresponds to the fluid (in the mixed phase on the ring) and
a maximum current phase which corresponds to the condensate on the ring.  
(We remind the reader that the densities in the fluid depend on $q$
and $\lambda$ Only and not on the density $p=m=\rho$).

\section*{Acknowledgement}
We would like to thank Oleg Zaborovsky for reading the
manuscript and for useful comments.
This research was supported in part by the National Science Foundation
under Grant No. PHY99-07949.
%

%
\end{document}